%% file: gmsdi.tex
\documentclass{article}
\usepackage{spconf,amsmath,graphicx,lipsum,amsfonts,bbm,booktabs,tikz,pgfplots}
\usepackage{cprotect}

\title{Generalized Multi-Source Inference for Text Conditioned Music Diffusion Models}
\name{Emilian Postolache$^{1}$, Giorgio Mariani$^{1}$, Luca Cosmo$^{2}$, Emmanouil Benetos$^3$, Emanuele Rodolà$^1$\vspace{-2mm}}

\address{$^1$Sapienza University of Rome  \hspace{10mm} $^2$Ca' Foscari University of Venice \\  $^3$Queen Mary University of London}

\begin{document}
\ninept
\maketitle
\begin{abstract} 
Multi-Source Diffusion Models (MSDM) allow for compositional musical generation tasks: generating a set of coherent sources, creating accompaniments, and performing source separation. Despite their versatility, they require estimating the joint distribution over the sources, necessitating pre-separated musical data, which is rarely available, and fixing the number and type of sources at training time. This paper generalizes MSDM to arbitrary time-domain diffusion models conditioned on text embeddings. These models do not require separated data as they are trained on mixtures, can parameterize an arbitrary number of sources, and allow for rich semantic control. We propose an inference procedure enabling the coherent generation of sources and accompaniments. Additionally, we adapt the Dirac separator of MSDM to perform source separation. We experiment with diffusion models trained on Slakh2100 and MTG-Jamendo, showcasing competitive generation and separation results in a relaxed data setting.
\end{abstract}
\begin{keywords}
Music Generation, Diffusion Models, Source Separation
\end{keywords}
\section{INTRODUCTION}
\label{sec:intro}

The task of musical generation has seen significant advancements recently, thanks to developments in generative models. The families of generative models showcasing state-of-the-art results are latent language models \cite{van2017neural} and (score-based) diffusion models \cite{song2019generative, ho2020denoising, song2020score}. Latent language models map a continuous-domain (time or spectral) signal to a sequence of discrete tokens and estimate a density over such sequences autoregessively \cite{agostinelli2023musiclm, copet2023simple} or via mask-modeling \cite{garcia2023vampnet}. 
Diffusion models \cite{schneider2023mo, liu2023audioldm}, on the other hand, operate on continuous representations (time, spectral, or latent domains), capturing the gradient of the log-density perturbed by a nosing process (Gaussian). Despite differences between these generative models, they typically share some mechanisms for conditioning on rich textual embeddings, obtained either using text-only encoders \cite{raffel2020exploring} or audio-text contrastive encoders \cite{manco2022learning, elizalde2023clap, wu2023large}. Such a mechanism allows generating a musical track following a natural language prompt. 

Generative models for music typically output only a final mixture. As such, generating the constituent sources is challenging. This implies that musical generative models are hard to employ in music production tasks, where the subsequent manipulation of sub-tracks, creation of accompaniments, and source separation is often required. 
Two existing approaches aim to address this issue. The first approach, called Multi-Source Diffusion Models (MSDM) \cite{mariani2023multi}, trains a diffusion model in time domain on (supervised) sets of coherent sources viewed as different channels without conditioning on textual information. Such a model allows for generating a set of coherent sources, creating accompaniments, and performing source separation. Despite being a versatile compositional model for music, MSDM has three limitations: (i) It requires knowledge of separated coherent sources, which are hard to acquire. (ii) It architecturally assumes a fixed number of sources and their respective class type (e.g., Bass, Drums, Guitar, Piano). (iii) It is impossible to condition the sources on rich semantic information, as commonly done with text-conditioned music models. The second approach, based on supervised instruction prompting \cite{wang2023audit, han2023instructme}, fine-tunes a latent diffusion model with instructions that allow adding, removing, and extracting sources present in a musical track. Although this approach addresses the issues (ii) and (iii) of MSDM, it does not solve the problem (i), necessitating pre-separated data. A strategy for scaling both models is training with data obtained by separating sources from mixtures using a pre-trained separator  \cite{donahue2023singsong}. This approach, though, is not flexible because such separated data contains artifacts, and we are limited to the number and type of sources the separator can handle.

We develop a novel inference procedure for the task, called \textit{Generalized Multi-Source Diffusion Inference (GMSDI)}, that can be used in combination with \textit{any} text-conditioned (time-domain) diffusion model for music. Such a method: (i) Requires only mixture data for training, resulting in an unsupervised algorithm when paired with a contrastive encoder. (ii) Parameterizes an arbitrary number and type of sources. (iii) Allows for rich semantic control. To our knowledge, this is the first general algorithm for unsupervised compositional music generation. After developing the required background notions in Section \ref{sec:background}, we develop the inference techniques in Section \ref{sec:gmsdi}. We detail the experimental setup in Section \ref{sec:setup} and show empirical results in Section \ref{sec:results}. We conclude the paper in Section \ref{sec:conclusion}.

\begin{figure*}[t!]
  \centering
  \centerline{\includegraphics[width=0.80\linewidth]{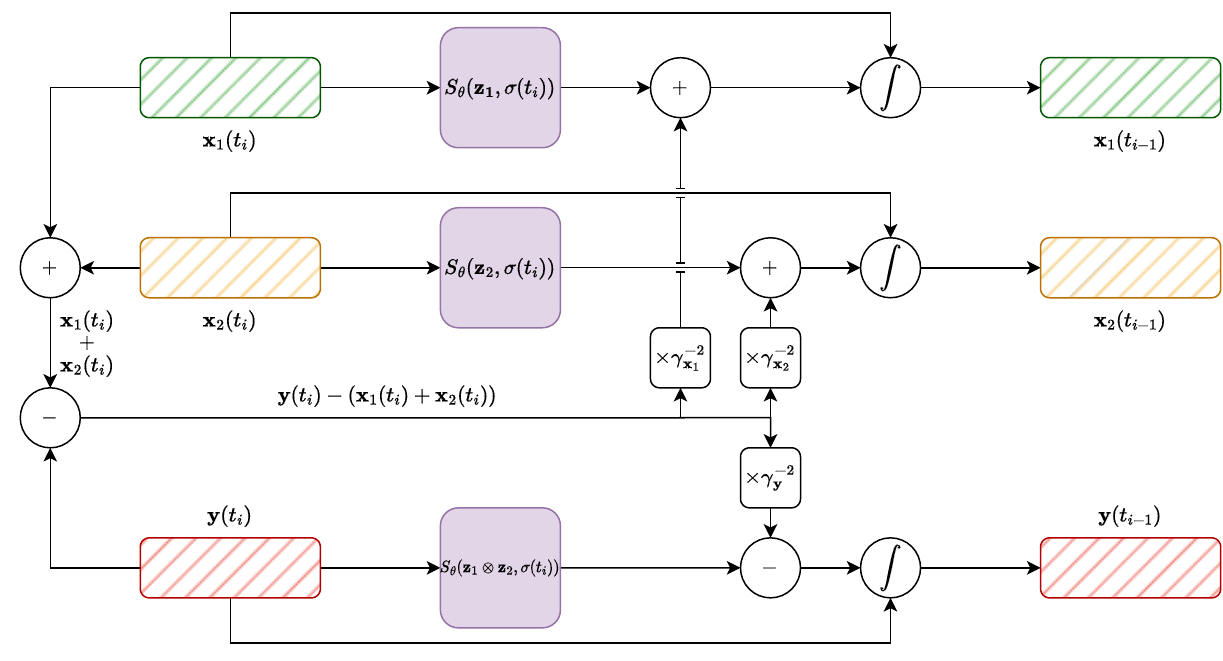}}
  \vspace{-0.2cm}
\caption{Diagram for unconditional generation procedure with GMSDI, sampling two coherent sources.}
\label{fig:gmsdi}
\end{figure*}

\section{BACKGROUND}
\label{sec:background}

A musical track $\mathbf{y}$ is a mixture of $K$  instrumental and vocal sources $\mathbf{x} = \{\mathbf{x}_k\}_{k \in [K]}$. Therefore, we have $\mathbf{y} = \sum_{k = 1}^{K} \mathbf{x}_k$, with $K$ depending on the mixture. Fixing a source $\mathbf{x}_k$, we denote the complementary set with $\mathbf{x}_{\bar{k}} = \{ \mathbf{x}_l\}_{l\in [K]} - \{\mathbf{x}_k\}$. While we typically do not have direct access to the audio constituents $\{\mathbf{x}_k\}_{k \in [K]}$, we are usually equipped with a text embedding $\mathbf{z}$ which provides information about the sources. We can obtain $\mathbf{z} = E^{\text{text}}_\phi(\mathbf{q})$ by  encoding a text description $\mathbf{q}$ with a text-only encoder $E^{\text{text}}_\phi$, or use a pre-trained contrastive audio-text encoder $E^{\text{contr}}_\phi$ to extract embeddings both from the audio mixtures $\mathbf{z} = E^{\text{contr}}_\phi(\mathbf{y})$ and from text descriptions $\mathbf{z} = E^{\text{contr}}_\phi(\mathbf{q})$.

\subsection{Text-conditioned Score-based Diffusion Models}
\label{subsec:score_based}
We work with continuous-time score-based \cite{song2020score} diffusion models. A text-conditioned score-based diffusion model $S_\theta$ parameterizes the logarithm of the perturbed audio mixture density, conditioned on the textual embedding:
\begin{equation}
\label{eq:text_diffusion}
\nabla_{\mathbf{y}(t)} \log p(\mathbf{y}(t) \mid \mathbf{z}) \approx S_\theta(\mathbf{y}(t), \mathbf{z}, \sigma(t))\,,
\end{equation}
where $p(\mathbf{y}(t) \mid \mathbf{z}) = \int_{\mathbf{y}(0)}p(\mathbf{y}(t) \mid \mathbf{y}(0))p(\mathbf{y}(0) \mid \mathbf{z})$, with
\begin{equation} 
	\label{eq:perturbation_kernel}p(\mathbf{y}(t) \mid  \mathbf{y}(0)) = \mathcal{N}(\mathbf{y}(t) \mid \mathbf{y}(0), \sigma^2(t) \mathbf{I})\,
\end{equation} a Gaussian perturbation kernel depending on a noise schedule $\{\sigma(t)\}_{t\in [0, T]}$. We train $S_\theta$ minimizing:
\begin{equation*}
\mathbb{E}_{t \sim \mathcal{U}[0,T]} \mathbb{E}_{
 \mathbf{y}(0) \sim p(\mathbf{y}(0) \mid \mathbf{z})} 
 \mathbb{E}_{\mathbf{y}(t) \sim p(\mathbf{y}(t) \mid \mathbf{y}(0))} \left[\mathcal{L}_{\text{SM}}\right]\,,
\end{equation*} where $\mathcal{L}_{\text{SM}}$ is the denoising score-matching loss \cite{song2019generative, ho2020denoising}:
\begin{equation*}
\mathcal{L}_{\text{SM}} = \Vert S_{\theta}(\mathbf{y}(t), \mathbf{z}, \sigma(t)) - \nabla_{\mathbf{y}(t)} \log p(\mathbf{y}(t) \mid \mathbf{y}(0))\Vert^2_2\,.
\end{equation*}
At inference time, we use classifier-free guidance \cite{ho2021classifierfree}, integrating
\begin{align*}
    &S^*_{\theta}(\mathbf{y}(t), \mathbf{z}, \sigma(t)) \\ =  \quad &S_{\theta}(\mathbf{y}(t), \mathbf{z}, \sigma(t)) + w (S_{\theta}(\mathbf{y}(t), \mathbf{z}, \sigma(t)) - S_{\theta}(\mathbf{y}(t), \mathbf{z}^*, \sigma(t)))\,,
\end{align*}
where $\mathbf{z}^*$ is a fixed learned embedding modeling the unconditional $\nabla_{\mathbf{y}(t)} \log p(\mathbf{y}(t))$, and $w \in \mathbb{R}$ is the embedding scale hyper-parameter. We can use a \textit{negative embedding} \cite{sanchez2023stay} instead of $\mathbf{z}^*$ to better guide inference. With an abuse of notation, we will refer to $S^*_\theta$ as $S_\theta$. 

\subsection{Multi-Source Diffusion Models}
In \cite{mariani2023multi}, authors assume a fixed number $K$ of coherent sources of known type $\{\mathbf{x}_{k}\}_{k\in [K]}$ contained in the mixture $\mathbf{y}$. They train a \textit{Multi-Source Diffusion Model (MSDM)}, an unconditional score-based diffusion model $S^{\text{MSDM}}_\theta$ that captures the joint distribution of coherent sources:
\begin{align}
\label{eq:msdm}
&\nabla_{(\mathbf{x}_1(t), \dots, \mathbf{x}_K(t))}\log p(\mathbf{x}_1(t), \dots, \mathbf{x}_K(t)) \nonumber \\ \approx  \quad  &S^{\text{MSDM}}_{\theta}((\mathbf{x}_1(t), \dots, \mathbf{x}_K(t)), \sigma(t))\,.
\end{align}
With the model, it is possible to perform music generation and source separation. \textit{Total (unconditional) generation} integrates Eq. \eqref{eq:msdm} directly,  generating all coherent sources  $\{\mathbf{x}_{k}\}_{k\in [K]}$ composing a track. \textit{Partial (conditional) generation} (i.e., accompaniment generation) fixes a known subset of sources $\mathbf{x}_{\mathcal{I}} = \{\mathbf{x}_i\}_{i \in \mathcal{I}}$ ($\mathcal{I} \subset [K]$) and generates the complementary subset $\mathbf{x}_{\bar{\mathcal{I}}}$ ($\bar{\mathcal{I}} = [K] - \mathcal{I}$) coherently. \textit{Source separation} extracts all sources from an observable mixture $\mathbf{y}(0)$, integrating, for all $k$, the approximate posteriors $\nabla_{\mathbf{x}_k(t)} \log(\mathbf{x}(t) \mid \mathbf{y}(0))$, modeled with Dirac delta likelihood functions. They propose a contextual separator using $S^{\text{MSDM}}_\theta$ and a weakly supervised separator, using a model $S_{\theta, k}$ for each source type. When constraining the last source, the weakly supervised separator samples from:
\begin{align}
\label{eq:weak_separator}
	S_{\theta,k}(\mathbf{x}_k(t), \sigma(t)) - S_{\theta, K}(\mathbf{y}(0) - \sum_{k=1}^{K-1}\mathbf{x}_{k}(t), \sigma(t))\,.
\end{align}

\section{GENERALIZED MULTI-SOURCE DIFFUSION INFERENCE}
\label{sec:gmsdi}
We train (or use) a text-conditioned diffusion model (Eq. \eqref{eq:text_diffusion}) $S_\theta(\mathbf{y}(t), \mathbf{z}, \sigma(t))$, with pairs of audio mixtures $\mathbf{y}(t)$ and associated text embeddings $\mathbf{z}$, containing information about the sources present in the mixture. We assume that each text embedding $\mathbf{z}$ is of the form $ \mathbf{z}_1 \otimes \dots \otimes \mathbf{z}_K$ (more compactly $\bigotimes_{k=1}^{K} \mathbf{z}_k$), where each $\mathbf{z}_k$ describes a source $\mathbf{x}_k$ present in $\mathbf{y}$ and $\otimes$ denotes an encoding of concatenated textual information (e.g., $\mathbf{z}_1 \otimes \dots \otimes \mathbf{z}_K = E_\phi^{\text{text}}(\mathbf{q}_1, \dots, \mathbf{q}_K)$, with $E_\phi^{\text{text}}(\mathbf{q}_k) = \mathbf{z}_k$). The idea is to leverage such text embeddings for parameterizing the individual source score functions:
\begin{equation}
\label{eq:source_diffusion}
 \nabla_{\mathbf{x}_k(t)} \log p(\mathbf{x}_k(t) \mid \mathbf{z}_k) \approx  S_\theta(\mathbf{x}_k(t), \mathbf{z}_k, \sigma(t))\,,
\end{equation} even if the model is trained only on mixtures.
We devise a set of inference procedures for $S_\theta$, called \textit{Generalized Multi-Source Diffusion Inference}, able to solve the tasks of $S^\text{MSDM}_\theta$ in the relaxed data setting.

\subsection{Total generation}

In order to generate a coherent set of sources $\{ \mathbf{x}_k\}_{k\in [K]}$, described by text embeddings $\{\mathbf{z}_k\}_{k \in [K]}$, we can sample from the conditionals $p(\mathbf{x}_k(t) \mid \mathbf{x}_{\bar{k}}(t), \mathbf{y}(t), \mathbf{z}_1, \dots, \mathbf{z}_K, \mathbf{z}_1 \otimes \dots \otimes \mathbf{z}_K)$:
\begin{equation}
\label{eq:joint}
\frac{p(\mathbf{x}(t), \mathbf{y}(t) \mid \mathbf{z}_1, \dots, \mathbf{z}_K, \mathbf{z}_1 \otimes \dots \otimes \mathbf{z}_K)}{p(\mathbf{x}_{\bar{k}}(t), \mathbf{y}(t) \mid \mathbf{z}_{\bar{k}}, \mathbf{z}_1 \otimes \dots \otimes \mathbf{z}_K)}\,.
\end{equation}
First, we develop the numerator in Eq. \eqref{eq:joint} using the chain rule:
\begin{align}
&p(\mathbf{x}(t),\mathbf{y}(t) \mid \mathbf{z}_1, \dots, \mathbf{z}_K, \mathbf{z}_1 \otimes \dots \otimes \mathbf{z}_K)  \nonumber \\ 
 = \quad &p(\mathbf{x}_k(t)\mid  \mathbf{z}_k)p(\mathbf{y}(t), \mathbf{x}_{\bar{k}}(t) \mid \mathbf{x}_k(t), \mathbf{z}_{\bar{k}}, \mathbf{z}_1 \otimes \dots \otimes \mathbf{z}_K)  \nonumber  \\
 = \quad &p(\mathbf{x}_k(t)\mid  \mathbf{z}_k)p(\mathbf{y}(t) \mid \mathbf{x}(t)) p(\mathbf{x}_{\bar{k}}(t) \mid \mathbf{x}_k(t), \mathbf{z}_{\bar{k}}) \nonumber \\
 \approx \quad &p(\mathbf{x}_k(t)\mid  \mathbf{z}_k)p(\mathbf{y}(t) \mid \mathbf{x}(t)) 
 \label{eq:conditional_xk}\,.
\end{align} 
We assume independence of the likelihood $p(\mathbf{y}(t) \mid \mathbf{x}(t))$ from embeddings and approximate the last equality dropping the unknown term $p(\mathbf{x}_{\bar{k}}(t) \mid \mathbf{x}(t), \mathbf{z}_{\bar{k}})$.
We substitute Eq. \eqref{eq:conditional_xk} in Eq. \eqref{eq:joint}, take the gradient of the logarithm with respect to $\mathbf{x}_k(t)$ and model the likelihood with isotropic Gaussians \cite{jayaram2020source} depending on a variance $\gamma_{\mathbf{x}_k}^2$:
\begin{align}
&\nabla_{\mathbf{x}_k(t)}\frac{\log p(\mathbf{x}_k(t)\mid  \mathbf{z}_k)p(\mathbf{y}(t) \mid \mathbf{x}(t))}{\log p(\mathbf{x}_{\bar{k}}(t), \mathbf{y}(t) \mid \mathbf{z}_{\bar{k}}, \mathbf{z}_1 \otimes \dots \otimes \mathbf{z}_K)} \nonumber \\ \nonumber 
= &\nabla_{\mathbf{x}_k(t)}\log p(\mathbf{x}_k(t) \mid \mathbf{z}_k) +  \nabla_{\mathbf{x}_k(t)}\log p(\mathbf{y}(t) \mid \mathbf{x}(t)) \\ \nonumber
= &\nabla_{\mathbf{x}_k(t)}\log p(\mathbf{x}_k(t) \mid \mathbf{z}_k) +  \nabla_{\mathbf{x}_k(t)} \log \mathcal{N}(\mathbf{y}(t) \mid \sum_{l = 1}^K \mathbf{x}_l(t), \gamma_{\mathbf{x}_k}^2\mathbf{I})\\ 
= &\nabla_{\mathbf{x}_k(t)}\log p(\mathbf{x}_k(t) \mid \mathbf{z}_k) + \frac{1}{\gamma_{\mathbf{x}_k}^2} (\mathbf{y}(t) - \sum_{l=1}^{K} \mathbf{x}_l(t))\,.\label{eq:gradient_x}  
\end{align}
Applying similar steps we obtain the score of the density on $\mathbf{y}(t)$ conditioned on $\mathbf{x}(t)$ (notice the opposite likelihood gradient):
\begin{align}
\nonumber
    &p(\mathbf{y}(t) \mid \mathbf{x}(t), \mathbf{z}_1, \dots, \mathbf{z}_K, \mathbf{z}_1 \otimes \dots \otimes \mathbf{z}_K)
    \\ \approx \quad 
&\nabla_{\mathbf{y}(t)} \log p(\mathbf{y}(t) \mid \bigotimes_{l=1}^{K} \mathbf{z}_l) + \frac{1}{\gamma_{\mathbf{y}}^2} (\sum_{l=1}^{K} \mathbf{x}_l(t) - \mathbf{y}(t))\,. 
\label{eq:gradient_y}  
\end{align}
During inference, we sample from Eqs. \eqref{eq:gradient_x} and \eqref{eq:gradient_y} in \textit{parallel}, replacing the gradients of the log-densities with score models (Eq. \eqref{eq:source_diffusion}):
\begin{equation}
\label{eq:gmsdi_score}
\begin{cases}
S_\theta(\mathbf{x}_k(t), \mathbf{z}_k, \sigma(t)) + \frac{1}{\gamma_{\mathbf{x}_k}^2} (\mathbf{y}(t) - \sum_{l=1}^{K} \mathbf{x}_l(t)) \\
  S_\theta(\mathbf{y}(t), \bigotimes_{l=1}^K \mathbf{z}_l, \sigma(t)) + \frac{1}{\gamma_{\mathbf{y}}^2} ( \sum_{l=1}^{K} \mathbf{x}_l(t) - \mathbf{y}(t))
\,.
\end{cases}
\end{equation}
A diagram of the method is illustrated in Figure \ref{fig:gmsdi}. Given a partition $\{\mathcal{J}_m\}_{m\in[M]}$ of $[K]$ containing $M$ subsets (i.e.,  $\cup_{m\in [M]}\mathcal{J}_m = [K]$), we can perform inference more generally with:
\begin{equation}
\label{eq:general_gmsdi_score}
\resizebox{1.\linewidth}{!}{%
$\begin{cases}
 S_\theta(\sum_{j \in \mathcal{J}_m}\mathbf{x}_j(t),  \bigotimes_{j \in \mathcal{J}_m} \mathbf{z}_j, \sigma(t)) + \frac{1}{\gamma_{\mathcal{J}_m}^2} (\mathbf{y}(t) - \sum_{l=1}^{K} \mathbf{x}_l(t)) 
 \\
  S_\theta(\mathbf{y}(t), \bigotimes_{l=1}^K \mathbf{z}_l, \sigma(t)) + \frac{1}{\gamma_{\mathbf{y}}^2} ( \sum_{l=1}^{K} \mathbf{x}_l(t) - \mathbf{y}(t)).
\end{cases}$
}%
\end{equation}
\subsection{Partial generation}
We can generate accompaniments $\mathbf{x}_\mathcal{J}$ for a given set of sources $\mathbf{x}_{\mathcal{I}}$, described by $\{\mathbf{z}_{i}\}_{i \in \mathcal{I}}$, by selecting a set of accompaniment text embeddings $\{\mathbf{z}_j\}_{j \in \mathcal{J}}$. We integrate Eqs. \eqref{eq:gmsdi_score} for $j \in \mathcal{J}$:
\begin{equation}
    \label{eq:gmsdi_partial}
    \resizebox{1.\linewidth}{!}{%
$\begin{cases}
 S_\theta(\mathbf{x}_j(t),  \mathbf{z}_j(t), \sigma(t)) + \frac{1}{\gamma_{\mathbf{x}_j}^2} \left[\mathbf{y}(t) - \left(\alpha \sum_{i\in \mathcal{I}} \mathbf{x}_i(t) + \beta \sum_{l\in \mathcal{J}} \mathbf{x}_l(t)\right)\right] 
 \\
  S_\theta(\mathbf{y}(t), \bigotimes_{l=1}^K \mathbf{z}_l, \sigma(t)) + \frac{1}{\gamma_{\mathbf{y}}^2} \left[\left( \alpha \sum_{i\in \mathcal{I}} \mathbf{x}_i(t) + \beta \sum_{l\in \mathcal{J}} \mathbf{x}_l(t)\right) - \mathbf{y}(t)\right]\,,
\end{cases}
$}%
\end{equation}
  with $\mathbf{x}_i(t)$ $(i \in \mathcal{I})$ sampled from the perturbation kernel in Eq. \eqref{eq:perturbation_kernel} conditioned on $\mathbf{x}_i$ and $\alpha, \beta \in \mathbb{R}$ scaling factors. Using Eq. \eqref{eq:general_gmsdi_score}, we can generate the accompaniment mixtures  $\sum_{j \in \mathcal{J}}\mathbf{x}_{j}$ directly.

\begin{figure}[t!]
  \centering
  \scalebox{0.95}{
    \input{Figures/fad_single.tex}}
 \caption{FAD (lower is better) between generated sources and Slakh100 test data (200 chunks, $\sim$12s each). Neg Prompt indicates the presence of negative prompting.}
	\label{fig:fad_single}
\end{figure}
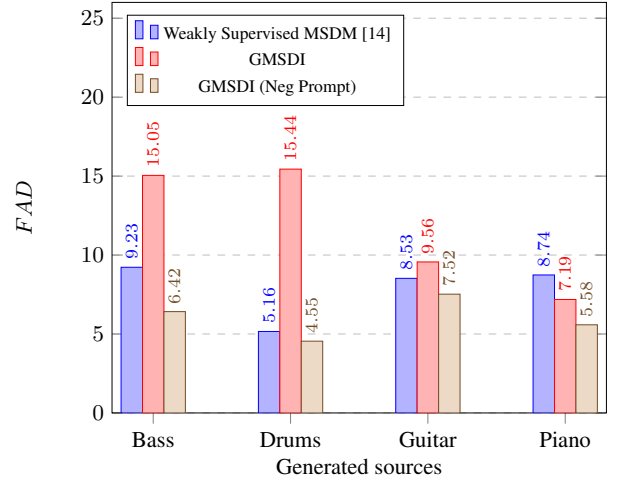

\subsection{Source separation}
Source separation can be performed by adapting Eq. \eqref{eq:weak_separator} to the text-conditioned model. Let an observable mixture $\mathbf{y}(0)$ be composed by sources described by $\{\mathbf{z}_k\}_{k\in [K]}$. We can separate the sources by choosing a constrained source (w.l.o.g. the $K$-th) and sampling, for $k \in [K-1]$, with:
\begin{equation}
\label{eq:gmsdi_separator}
    S_{\theta}(\mathbf{x}_k(t), \mathbf{z}_k, \sigma(t)) - S_{\theta}(\mathbf{y}(0) - \sum_{l=1}^{K-1}\mathbf{x}_{l}(t), \mathbf{z}_K, \sigma(t))\,.
\end{equation} We call this method \textit{GMSDI Separator}. We also define a \textit{GMSDI Extractor}, where we extract the $k$-th source $\mathbf{x}_{k}$ with:
\begin{equation}
\label{eq:i_gmsdi_extractor}
    	S_{\theta}(\mathbf{x}_k(t),  \mathbf{z}_k, \sigma(t)) - S_{\theta}(\mathbf{y}(0) - \mathbf{x}_k(t),  \bigotimes_{l \neq k} \mathbf{z}_l, \sigma(t))\,,
\end{equation}
constraining the  mixture $\sum_{l\neq k} \mathbf{x}_l(t)$, complementary to $\mathbf{x}_k(t)$.

\section{EXPERIMENTAL SETUP}
\label{sec:setup}
To validate our theoretical claims, we train two time-domain Moûsai-like \cite{schneider2023mo} diffusion models. The first model is trained on Slakh2100 \cite{manilow2019cutting}. Slakh2100 is a dataset used in source separation, containing 2100 multi-source waveform music tracks obtained by synthesizing MIDI tracks with high-quality virtual instruments. We train the diffusion model on mixtures containing the stems Bass, Drums, Guitar, and Piano (the most abundant classes). To condition the diffusion model, we use the \verb|t5-small| pre-trained T5 text-only encoder \cite{raffel2020exploring}, which inputs the concatenation of the stem labels present in the mixture (e.g., ``Bass, Drums" if the track contains Bass and Drums). Given that we know the labels describing the sources inside a mixture at training time, such an approach is weakly supervised. The  window size is $2^{18}$ at 22kHz ($\sim$12s). 

The second model is trained on a more realistic dataset, namely MTG-Jamendo \cite{bogdanov2019the}. MTG-Jamendo is a music tagging dataset containing over 55000 musical mixtures and 195 tag categories. We train our diffusion model on the \verb|raw_30s/audio-low| version of the dataset, using the first 98 shards for training and the last 2 for validation. The model window is of $2^{19}$ samples ($\sim$24s) at 22kHz.  We condition the model with the pre-trained checkpoint \verb|music_audioset_epoch_15_esc_90.14.pt|\cprotect\footnote{\verb|https://github.com/LAION-AI/CLAP|} of the LAION CLAP  contrastive encoder  \cite{wu2023large}. At training time, we condition the diffusion model with embeddings $E_\phi^{\text{contr}}(\mathbf{y})$ obtained from the training mixtures $\mathbf{y}$ themselves, resulting in an unsupervised model. At inference time, we use ADPM2\cprotect\footnote{\verb|https://github.com/crowsonkb/k-diffusion|} \cite{lu2022dpm} with $\rho = 1$ for generation and AEuler$^\text{2}$ with  $s_\text{churn} = 20$ for separation.

\section{EXPERIMENTAL RESULTS}
\label{sec:results}

\begin{figure}[t!]
  \centering
 \scalebox{0.95}{
    \input{Figures/fad_gen.tex}}
 \caption{FAD (lower is better) results on total and partial generation, with respect to Slakh2100 test mixtures (200 chunks, $\sim$12s each).}
 \label{fig:fad_gen}
\end{figure}
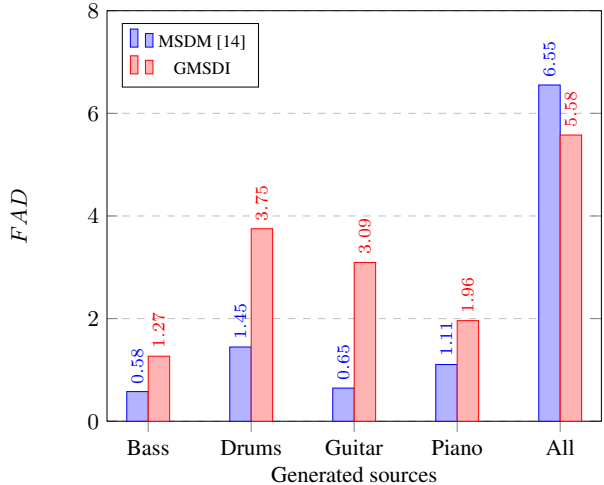

First, we want to understand whether the model trained on Slakh2100 mixtures can parameterize single sources well. We sample, for each stem, 200 chunks of $\sim$12s, conditioning with embeddings of single stem labels (e.g., ``Bass"). Then, we compute the Fréchet Audio Distance (FAD) \cite{kilgour2019frechet} with VGGish embeddings between such samples and 200 random Slakh2100 test chunks of the same source. In Figure \ref{fig:fad_single}, we compare our model against the weakly supervised version of MSDM  \cite{mariani2023multi}, where a model learns the score function for each stem class (a setting requiring access to clean sources). We notice that single-stem prompting is insufficient for obtaining good FAD results, especially for Bass and Drums, causing silence to be generated. We find negative prompts (Section \ref{subsec:score_based}) essential for obtaining non-silent results using ``Drums, Guitar, Piano" (Bass),  ``Bass" (Drums),  ``Bass, Drums" (Guitar), ``Bass, Drums" (Piano). In all settings above, we use 150 sampling steps.

Following, we ask how well the model can perform coherent synthesis with GMSDI. In Figure \ref{fig:fad_gen}, we compute the FAD between 200 random Slakh2100 test mixture chunks ($\sim$ 12s each) and mixture chunks obtained by summing the model's generated stems (unconditional) or the generated stems together with the conditioning tracks (conditional). On total generation (All), we set $\gamma_{\mathbf{y}} = \infty$ and reach $\sim$ 1 lower FAD point, using 600 sampling steps. On partial generation, we sample using 300 steps, setting $\gamma_{\mathbf{y}} \ll \infty$, to inform the generated mixture about the conditioning sources. In this scenario, MSDM tends to generate silence. To enforce non-silent results with MSDM, we sample 100 examples for each conditioning chunk and select the sample with the highest $L_2$ norm.

For source separation, we employ the SI-SDR improvement (SI-SDR$_{\text{i}}$) \cite{le2019sdr} as an evaluation metric and follow the evaluation protocol of \cite{mariani2023multi}. First, we perform a grid search (Table \ref{tab:results_grid}) to find a good embedding scale $w$. For the GMSDI Separator, we do not use negative prompting, while for the GMSDI Extractor, we only use negative prompts for Bass and Drums. We evaluate on the full Slakh2100 test set with $w = 3$ and constrained Drums for GMSDI Separator and $w = 7.5$ for GMSDI Extractor, showcasing results in Table \ref{tab:results_sep}. Training only with mixtures (plus associated labels), the ensemble of the two separators reaches 11.56 dB, being zero-shot, i.e., we do not target source separation during training \cite{pons2023gass}. 

We release qualitative examples for the Slakh2100 and MTG-Jamendo models on our demo page\cprotect\footnote{\verb|https://github.com/gladia-research-group/gmsdi|}.

\begin{table}[t]
\centering
\caption{Grid search over embedding scale $w$ on 100 chunks ($\sim$12s each) of Slakh2100 test set. Results in $\text{SI-SDR}_\text{i}$ (dB -- higher is better). The source in parenthesis is the constrained source.}
\label{tab:results_grid}
\resizebox{1.\linewidth}{!}{%
 \begin{tabular}{l  c c c c} 
 \toprule
 \textbf{Model}  & $w = 3.0$ &  $w = 7.5$ & $w = 15.0$ & $w = 24.0$ \\ 
 \midrule
   GMSDI Extractor & 7.66 & \textbf{9.61} & 6.00 & -0.62 \\
  GMSDI Separator (Bass) & 8.10 & 6.72 & -1.09 & -20.60 \\
    GMSDI Separator (Drums) & \textbf{9.44} & 8.69 & -1.48 & -21.62 \\
  GMSDI Separator (Guitar) & 5.82 & 4.37 & -2.27 & -17.49 \\
  GMSDI Separator (Piano) & 7.60 & 6.41 & -2.68 & -16.90
  \\
 \bottomrule
\end{tabular}}%
\end{table}

\begin{table}[t]
\centering
\caption{Quantitative results for source separation on the Slakh2100 test set. Results in $\text{SI-SDR}_\text{i}$ (dB -- higher is better).}
\label{tab:results_sep}
\resizebox{1.\linewidth}{!}{%
 \begin{tabular}{l  c c c c  c} 
 \toprule
 \textbf{Model}  &\textbf{Bass} & \textbf{Drums} & \textbf{Guitar} & \textbf{Piano} & \textbf{All}\\ 
 \midrule
  Demucs + Gibbs (512 steps) \cite{ manilow2022improving}& 17.16 & 19.61 & 17.82 & 16.32 & \textbf{17.73} \\ 
  Weakly Supervised MSDM \cite{mariani2023multi} & 19.36 & 20.90 & 14.70 & 14.13 & 17.27 \\ 
  MSDM \cite{mariani2023multi} & 17.12 & 18.68 & 15.38 & 14.73 & 16.48 \\ \midrule
  GMSDI Separator & 9.76 & 15.57 & 9.13 &	9.57 & 11.01
 \\
  GMSDI Extractor & 11.00 & 10.55 & 9.52 & 10.13 & 10.30 \\
  Ensamble & 11.00 & 15.57 & 9.52 & 10.13 & \textbf{11.56} \\
 \bottomrule
\end{tabular}%
}
\end{table}

\section{CONCLUSIONS}
\label{sec:conclusion}
We have proposed GMSDI, a compositional music generation method working with any time-domain text-guided diffusion model. The method obtains reasonable generation and separation metrics on Slakh2100, enabling unsupervised compositional music generation for the first time. In future work, we want to extend the technique to latent diffusion models and narrow the gap with supervised methods.

\section{ACKNOWLEDGEMENTS}
This work is supported
by the ERC Grant no.802554 (SPECGEO) and PRIN 2020 project no.2020TA3K9N (LEGO.AI).
L.C. is supported by the IRIDE grant from DAIS, Ca’ Foscari University of Venice.
E.B. is supported by a RAEng/Leverhulme Trust Research Fellowship [grant no. LTRF2223-19-106].

\bibliographystyle{IEEEbib}
\bibliography{refs}

\end{document}

%% file: Figures/fad_single.tex
\begin{tikzpicture}
    \begin{axis}[
        ybar, 
        every node near coord/.append style={rotate=90, anchor=west},
        ymin=0, 
        ymax=26,
        symbolic x coords={Bass, Drums, Guitar, Piano}, 
         xlabel={Generated sources},
        ylabel={$FAD$},
           ybar=0.0cm, 
        bar width=0.3cm, 
        xtick=data,                     
        xticklabel pos=lower,           
        every node near coord/.append style={
            font=\scriptsize
        },
        xtick pos=bottom,   
        ymajorgrids=true,
        grid style=dashed,
        legend pos=north west,
        nodes near coords,
        width=8.5cm
    ]
    \addplot coordinates {(Bass, 9.226348703003659) (Drums,5.159943086886206) (Guitar,8.52502372116005) (Piano,8.738518639387056)};
    \addplot coordinates {(Bass,15.045405966323575) (Drums, 15.442264915619377) (Guitar, 9.56416320598704) (Piano,  7.1856969710831)};
    \addplot coordinates {(Bass, 6.4181531422464495) (Drums, 4.5459882669369005) (Guitar, 7.522703908687857) (Piano, 5.583175055879252)};
    \legend{\scriptsize Weakly Supervised MSDM \cite{mariani2023multi}, \scriptsize GMSDI, \scriptsize GMSDI (Neg Prompt)}
    \end{axis}
\end{tikzpicture}

%% file: Figures/fad_gen.tex
\begin{tikzpicture}
    \begin{axis}[
        ybar, 
        every node near coord/.append style={rotate=90, anchor=west},
        ymin=0,
        ymax=8,
        symbolic x coords={Bass, Drums, Guitar, Piano, All}, 
         xlabel={Generated sources},
        ylabel={$FAD$},
           ybar=0.0cm, 
        bar width=0.3cm, 
        xtick=data,                     
        xticklabel pos=lower,           
        every node near coord/.append style={
            font=\scriptsize
        },        
        xtick pos=bottom,   
        ymajorgrids=true,
        grid style=dashed,
        legend pos=north west,
        nodes near coords,
        nodes near coords style={/pgf/number format/precision=2},
        width=8.5cm
    ]

    \addplot coordinates {(Bass, 0.577980205712322) (Drums, 1.4463828746653444) (Guitar, 0.6464319420006781) (Piano,  1.1055633229768063) (All, 6.552404908666919 )};
    \addplot coordinates {(Bass, 1.2673057884589767) (Drums,3.7506752952650153) (Guitar,  3.0941195350071258) (Piano, 1.960139594906746) (All, 5.577163944251142)};
    
    \legend{ 
    \scriptsize MSDM \cite{mariani2023multi}, 
    \scriptsize GMSDI}
    \end{axis}
\end{tikzpicture}